\documentclass{ws-procs9x6}
\newcommand{\panda}{$\overline{\text{P}}$ANDA }

\begin{document}

\title{Development of a GEM-TPC prototype\\}

\author{Heinz Angerer$^1$, Reinhard Beck$^3$, Martin Berger$^*$, Felix B\"ohmer$^1$\\
  K. -T. Brinkmann$^3$, Paul B\"uhler$^4$, Michael Carnegie$^4$, Sverre D\o rheim$^1$, \\
  Laura Fabbietti$^5$, Chr. Funke$^3$, F. Cusanno$^5$, J\"org Hehner$^2$, Andreas Heinz$^2$, \\
  Markus Henske$^2$, Christian H\"oppner$^1$, David Kaiser$^3$, Bernhard Ketzer$^1$, \\
  Igor Konorov$^1$, Jochen Kunkel$^2$, Michael Lang$^3$, Johann Marton$^4$, \\
  Sebastian Neubert$^1$, Stephan Paul$^1$, Alexander Schmah$^5$, Christian Schmidt$^2$, \\
  Roman Schmitz$^3$, Sandra Schwab$^2$, Daniel Soyk$^2$, Ken Suzuki$^4$, Ulrike Thoma$^3$, \\
  Maxence Vandenbroucke$^1$, Bernd Voss$^2$, Dieter Walter$^3$, Quirin Weitzel$^1$, \\
  Eberhard Widmann$^4$, Alexander Winnebeck$^3$, Lisa W\"orner$^1$, H. -G. Zaunick$^3$, \\
  Xiaodong Zhang$^1$ and Johann Zmeskal$^4$ \\
  1 Technische Universit\"at M\"unchen, Physik Department E18, Garching, Germany\\
  2 Gesellschaft f\"ur Schwerionenforschung mbH, Darmstadt, Germany\\
  3 Helmholtz-Institut f\"ur Strahlen- und Kernphysik, Bonn, Germany\\
  4 Stefan-Meyer-Institut f\"ur subatomare Physik, Vienna, Austria\\
  5 Technische Universit\"at M\"unchen, Excellence Cluster ``Universe'', Garching, Germany}

\address{Technische Universit\"at M\"unchen, Excellence Cluster ``Universe'',\\
Garching, Germany 85748, Bavaria\\
$^*$E-mail: mberger@ph.tum.de\\
www.universe-cluster.de}



\begin{abstract}
The use of GEM foils for the amplification stage of a TPC instead of a conventional MWPC allows one to bypass the necessity of gating, as the backdrift is suppressed thanks to the asymmetric field configuration. 
This way, a novel continuously running TPC, which represents one option for the \panda  central tracker, can be realized. A medium sized prototype with a diameter of 300\,mm and a length of 600\,mm
will be tested inside the FOPI spectrometer at GSI using a carbon or lithium beam at intermediate energies (E\,=\,1-3AGeV). 
This detector test under realistic experimental conditions should allow us to verify the spatial resolution for single tracks and the reconstruction capability for displaced vertexes. A series of physics measurement implying pion beams is scheduled with the FOPI spectrometer together with the GEM-TPC as well.   
\end{abstract}


\bodymatter

\section{Introduction}\label{introduction}
To face the challenges of the physics program in $\overline{\text{P}}$ANDA (\underline{A}nti\underline{p}roton \underline{An}nihilations at \underline{Da}rmstadt), the cylindrical central tracker of the Target Spectrometer (TS) has to fulfill the following requirements\cite{panda2}\!: multiple track identification (up to 1000 tracks superimposed inside the TPC all the time), high spatial resolution ($\sigma_{r\varphi}\approx150\,\mu m,\sigma_z\approx1\,mm$), high momentum resolution ($\approx 1\,\%$), minimal material budget ($\approx 1\,\%$  of radiation length), high rate capability, resistance against aging, etc. Due to the beam characteristics, the TPC has to work in a continuous mode, i.e. without gating, which is another big challenge from the technical point of view. \\
A Time Projecting Chamber (TPC) with Gas Electron Multiplier (GEM) Readout not only fulfills all the requirements above, it furthermore provides very good dE/dx measurements also in the region of low momenta, which is very useful for particle identification.\\
In addition, the successful commissioning of the GEM-Detectors for the COMPASS (\underline{Co}mmon \underline{M}uon and \underline{P}roton \underline{A}pparatus for \underline{S}tructure and \underline{S}pectroscopy) experiment\cite{gem1} shows, that this kind of detector has excellent properties concerning high rate capability and intrinsic suppression of ion backflow. Such properties enables a GEM based TPC to operate in an ungated mode. In the next sections the actual development state, the progress and test results of a GEM-TPC prototype developed by our group will be discussed in detail.

\section{Test results of 80\,mm test chamber}
\begin{figure}[htb]
\centering
\psfig{file=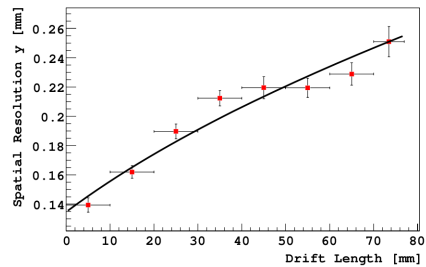,width=.5\textwidth}
\caption{The y-Axis resolution in dependency of the drift length. Red squares indicate the resolution along the y-Axis (short side of the pads). The black curve is a fit, taking diffusion into account.}
\label{pic_test_res}
\end{figure}

A first 80\,mm GEM-TPC test chamber has been build and tested in the laboratory with PASA/ALTRO \cite{pasa} (10\,MHz sampling rate) electronics using cosmic muons. It had a rectangular readout padplane with a pad size of 0.8\,mm\,x\,1.0\,mm and a pitch of 1.0\,mm\,x\,1.2\,mm. For details see Ref.\cite{gem4}. Here only the spatial resolution along the short side of the rectangular pads are shown (see Fig.~\ref{pic_test_res}).
From this figure one can see that the average spatial resolution along the short side of the pads is about 200$\mu$\,m. The Gas mixture used for this experiment was  Ar/CO$_2$ (70/30).
\section{Upgrade of the test chamber}
Since the charge distribution of rectangular pads is not homogeneous, due to the different distances between diagonal and direct neighboring pads, it has been decided to use hexagonal pads for the readout padplane. Simulations have shown that the spatial resolution saturates for a pad-radius smaller than 1.5mm, due mainly to diffusion effects. To validate this result, the new padplane for the test chamber has been equipped with pads in two different sizes, namely 1.5\,mm and 1.25\,mm. Furthermore the PASA/ALTRO based readout has been replaced by an AFTER based readout. This AFTER chip can be operated with 20\,MHz sampling rate and has much lower noise ($\approx 800\,e^-$) than the PASA/ALTRO chip ($\approx 1900\,e^-$).

\section{Test setup at ELSA}
\begin{figure}[htb]
\centering
\psfig{file=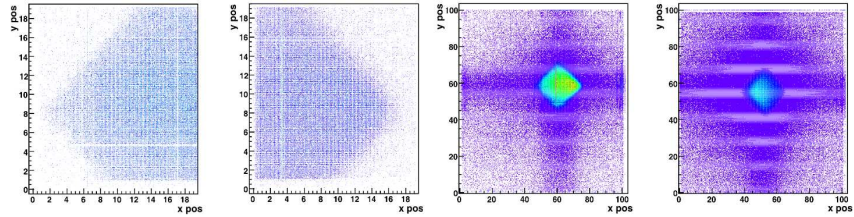,width=\textwidth}
\caption{Cluster hits of silicon strip detectors (left) and GEM detectors (right). Since the GEM2 is mounted downstream behind the TPC, the broadening of the distribution can be explained by multiple scattering. The geometry of the distributions is caused by the trigger scintillators.} 
\label{pic_bonn_beam}
\end{figure}

A beam telescope was set up at the ELSA accelerator facility to perform a first test of the GEM-TPC prototype with a low intensity electron beam ($E_beam\approx500\,MeV$) This telescope consists of two GEM-Detectors, four silicon strip detectors and four scintillators for triggering. During the first tests this beam telescope has been commissioned. The spatial resolutions of the reference detectors, which can be achieved at the moment, are shown in table \ref{tab_res}. The measured beam profile can be seen in Fig.~\ref{pic_bonn_beam}.

\begin{table}
 \centering
 \begin{tabular}{|c|c|c|c|c|}
 \hline
 Detector&GEM 1& GEM 2&Silicon 1&Silicon 2 \\\hline
 Y-Resolution [$\mu m$]&100&150&5&5\\\hline
 X-Resolution [$\mu m$]&200&250&10&10\\\hline
 \end{tabular}
\label{tab_res}
\end{table}

\section{First tests}
\begin{figure}[htb]
\centering
\psfig{file=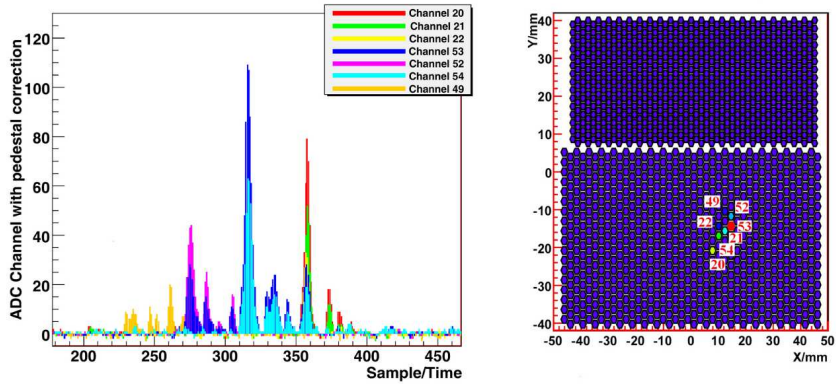,width=\textwidth}
\caption{On the left side the time structure of signals inside the test chamber are shown. All amplitudes are pedestal corrected. On the right side the corresponding pads are shown.} 
\label{pic_bonn_track1}
\end{figure}

\begin{figure}[htb]
\centering
\psfig{file=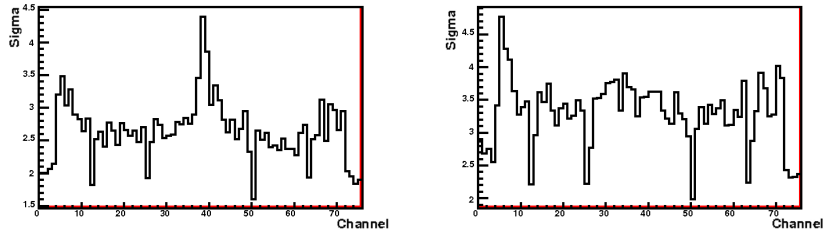,width=\textwidth}
\caption{Exemplary the sigma of the noise distribution for chips 10 and 11 versus channel number are shown. The peaks in the noise distributions are due to the track length and track position on the readout PCB. The dips are unconnected channels. In the future new frontend cards with shorter tracks will be used.} 
\label{pic_tpc_noise}
\end{figure}

During the first test it was possible to obtain first electron tracks inside the test chamber (see Fig.~\ref{pic_bonn_track1}).\\
For a second test the firmware of the readout electronics was updated and a full 2D beam profile was observed. Besides, the noise of the electronics could be reduced to a sigma of few adc channels (see Fig.~\ref{pic_tpc_noise}) The peaks in the noise distribution are due to the geometry of the tracks on the front-end card. For the main prototype the front end cards will get a different geometry to reduce this effects. The dips are caused by unconnected channels. Furthermore an alignment of the hole setup with photogrametry was done.

\section{GEM-TPC Prototype}
In parallel to the test with the small chamber, the construction of the prototype is ongoing. The first version of the padplane for the prototype has been finished and the connectors for the frontend cards were soldered on this padplane. A cooling device, the so called cooling-pot, has been machined and tested. Furthermore a flange to distribute the gas, the low and the high voltage was build. The high voltage testing of the GEM foils is finished successfully and some of the foils are already glued onto frames to build the GEM stack.

\section{Summary and Outlook}
The beam telescope was commissioned successfully and tracks could be reconstructed. In the tests with the test chamber first tracks and also a beam profile was obtained. Lots of data for further analysis were taken. The construction of the prototype is ongoing and many parts are already produced and tested.\\
In April 2010 a test of the prototype at the FOPI spectrometer at the ``Gesellschaft f\"ur Schwerionenforschung'' (GSI) with a proton beam with 3.0\,GeV kinetic beam energy is foreseen.

\bibliographystyle{ws-procs9x6}
\bibliography{lit_article}

\end{document}